\documentclass[acmsmall,screen,authorversion,nonacm]{acmart}
\usepackage{fancyhdr}

\AtBeginDocument{%
  \providecommand\BibTeX{{%
    \normalfont B\kern-0.5em{\scshape i\kern-0.25em b}\kern-0.8em\TeX}}
    
    \addtolength{\footskip}{2.0\baselineskip}%
    \fancyfoot[L]{\textit{\textbf{This is a preprint and has not undergone peer-review.}}}%
    
    }

\begin{document}

\title[Re-Ranking News Comments by Constructiveness, Curiosity, and Personal Stories]{Re-Ranking News Comments by Constructiveness and Curiosity Significantly Increases Perceived Respect, Trustworthiness, and Interest}

\author{Emily Saltz}
\affiliation{%
  \institution{Google Jigsaw}
  \country{USA}
  }
\email{emilysaltz@google.com}

\author{Zaria Jalan}
\affiliation{%
  \institution{Google Jigsaw}
  \country{USA}
  }
\email{zaria@google.com}

\author{Tin Acosta}
\affiliation{%
  \institution{Google Jigsaw}
  \country{USA}
  }
\email{tinacosta@google.com}







\renewcommand{\shortauthors}{Saltz et al.}

\begin{abstract}
Online commenting platforms have commonly developed systems to address online harms by removing and down-ranking content. An alternative, under-explored approach is to focus on up-ranking content to proactively prioritize prosocial commentary and set better conversational norms. We present a study with 460 English-speaking US-based news readers to understand the effects of re-ranking comments by constructiveness, curiosity, and personal stories on a variety of outcomes related to willingness to  participate and engage, as well as perceived credibility and polarization in a comment section. In our rich-media survey experiment, participants across these four ranking conditions and a control group reviewed prototypes of comment sections of a Politics op-ed and Dining article. We found that outcomes varied significantly by article type. Up-ranking curiosity and constructiveness improved a number of measures for the Politics article, including perceived \textit{Respect}, \textit{Trustworthiness}, and \textit{Interestingness} of the comment section. Constructiveness also increased perceptions that the comments were favorable to Republicans, with no condition worsening perceptions of partisans. Additionally, in the Dining article, personal stories and constructiveness rankings significantly improved the perceived informativeness of the comments. Overall, these findings indicate that incorporating prosocial qualities of speech into ranking could be a promising approach to promote healthier, less polarized dialogue in online comment sections.
\end{abstract}







\maketitle

\section{Introduction}

As social media and other user-generated content (UGC) platforms have grown to host more diverse users, the reigning content moderation paradigm has focused on reducing or removing the “bad stuff” across these users \cite{roberts2019behind,gillespie2022not}. This approach has been reasonable, as hate, harassment, and other toxic content can not only lead to real-world harms such as violent extremism \cite{binder2022terrorism} and self-harm \cite{martinez2020relationship}, but they can also drive away other users (and advertisers) from participating in those spaces in the first place \cite{smirnov2023toxic,IBMEUand66:online}. 
Yet even as hateful content is removed or downranked at an unprecedented scale \cite{YouTubeC12:online,TikTokre96:online,Communit78:online}, there is evidence that people are less satisfied with the state of online speech — and platforms’ moderation of it — than ever \cite{vogels2022support,saltz2021misinformation}. For news platforms in particular, many choose to close their comment sections entirely rather than moderate deeply polarized and toxic conversations \cite{nelson2021killing}. 

In this work, we investigate yet another technique: flipping the “bad stuff" paradigm on its head to promote prosocial content by re-ranking comments based on attributes associated with healthy dialogue and prosocial conversation. By focusing on re-ranking on prosocial attributes, such as constructiveness, curiosity, and personal stories (described in Section 2), it is possible to align ranking with high-level values for prosocial conversations without reliance on either sparse user survey data, or on behavioral signals like 'likes,' which provide an incomplete and crude representation of a user’s ideals.  

To understand the impact of such re-ranking, we focus on two primary questions:
\begin{itemize}
    \item \textbf{RQ1}. How does re-ranking by the attributes of constructiveness, curiosity, and personal stories affect readers’ perceptions of the quality of a news comment section?
    \item \textbf{RQ2}. How does the subject-matter of a news comment section affect readers’ perceptions of quality of comments re-ranked by constructiveness, curiosity, and personal stories? 
    
\end{itemize} We measured quality using outcome categories of \textit{Participation}, \textit{Credibility}, \textit{Engagement}, and \textit{Polarization}, described in Section 3.3. 

\section{Related Work and Prosocial Attributes}
There have been a number of efforts\footnote{Other examples: MIT’s Center for Constructive Communication Lab and Cortico, who develop tools to find “bridging” points of commonality in live conversations \cite{Overview19:online}. A proposal to align recommender systems with human values dynamically through community-level well-being metrics \cite{stray2020aligning}. The Prosocial Design Network hosts a collection of other prosocial intervention examples \cite{Prosocia63:online}. 
} in recent years exploring interventions that promote prosocial elements in online conversations. For example, platforms like X (formerly Twitter) and Polis \cite{megill2022coherent}, have used “bridging” algorithms to promote content that has support across users who have formerly disagreed with each other, with varying success \cite{allen2022birds}. YouTube has also championed approaches to “raise” and “reward” authoritative content \cite{thefourrs}, and has long incorporated signals from user surveys to align with viewer satisfaction \cite{covington2016deep}.

Which attributes of online speech are most closely tied to prosocial outcomes? The study of conflict \cite{kugler2011moral}, as well as linguistic markers of conversations, have also identified a plethora of prosocial qualities present in online comment corpora \cite{kolhatkar2020classifying,choi2020ten,10.1145/3442381.3450122}. 

Drawing on this body of literature, we developed three experimental classifiers\footnote{While we explored other attributes such as compassion, nuance, similarity, and respect, these above attributes were chosen for focus in this study as they were the highest performing in AUC-ROC evaluations at the time of this study. Model card with more details included in the Appendix C.} for attributes associated with prosocial behavior online, defined in Table 1. In this study, we focus on how re-ranking by these constructiveness, curiosity, and personal stories in news comment sections affects different prosocial outcomes compared to control rankings. We chose ranking because it is a prominent design feature of many large online platforms, and thus one that might be particularly meaningful to manipulate. We chose news comments as they tend to be particularly fraught and polarized, in which vital civic information and political knowledge is shared and discussed.

\begin{table}
  \caption{Three attributes alongside definitions, supporting evidence, and example comments. Full comments in Appendix A. stimuli.}
  \label{tab:prosociality}
  \footnotesize
  \begin{tabular}{p{2.5cm}p{3.5cm}p{3.5cm}p{3cm}}
    \toprule
    Attribute & Definition & Evidence of Prosociality & Example Comments \\
    \midrule
    Constructiveness & Makes specific or well-reasoned points to provide a fuller understanding of the topic without disrespect or provocation. & ‘Specific points’ and ‘evidence’ \cite{kolhatkar2020classifying}; ‘integrative complexity,’ to incorporate multiple perspectives into reasoning \cite{kugler2011moral} & \textit{“That the criticism of this piece includes references to ‘both-sider-ism’ is understandable...”} \\
    \addlinespace
    Curiosity & Attempts to clarify or ask follow-up questions to better understand another person or topic. & ‘Dialogue’ \cite{kolhatkar2020classifying}; ‘behavioral complexity’: in which people explicitly ask questions designed to better understand a point of view \cite{kugler2011moral} & \textit{“In these surveys, are Democrats asked what they mean by ‘Republicans’...”} \\
    \addlinespace
    Personal Stories & Shares a personal experience or story or that of someone the author knows within the comment. & ‘Personal stories’ \cite{kolhatkar2020classifying}; ‘Personal disclosure’ \cite{10.1145/3442381.3450122}. & \textit{“Four years ago I was eating dinner at the home of a lifelong friend, a conservative man...”} \\
    \bottomrule
  \end{tabular}
\end{table}

\section{Methodology}
\subsection{Participants and recruiting}
We recruited news readers from user research platform dscout \cite{dscoutFl6:online} to participate in a rich-media survey experiment using its “Express Media Survey” format \cite{Whatisds68:online}, which enables deployment of survey questions alongside unmoderated video screen-recordings of prototypes. We initially recruited 500 US-based, English-speaking participants, or 125 per group. All participants were compensated \$15. All self-reported being news readers, with a mix of commenting habits. We applied a demographic quotas to ensure diversity so that no age, ethnicity, or gender would exceed 65\% amongst respondents. 
We removed 40 entries total due to data quality issues.\footnote{Entries were removed if a participant applied and completed the survey for more than one condition, in which cases, the first entry was preserved. Entries were also removed if participants completed the survey in an unreasonably fast time period, e.g. under five minutes, or if participants reported being unable to open the prototype link.} The final sample size was 460, with slight variations in participant counts per group, as described in Figure 1. 

\subsection{Experiment design and stimuli}
\begin{figure}[h]
  \centering
  \includegraphics[width=\textwidth]{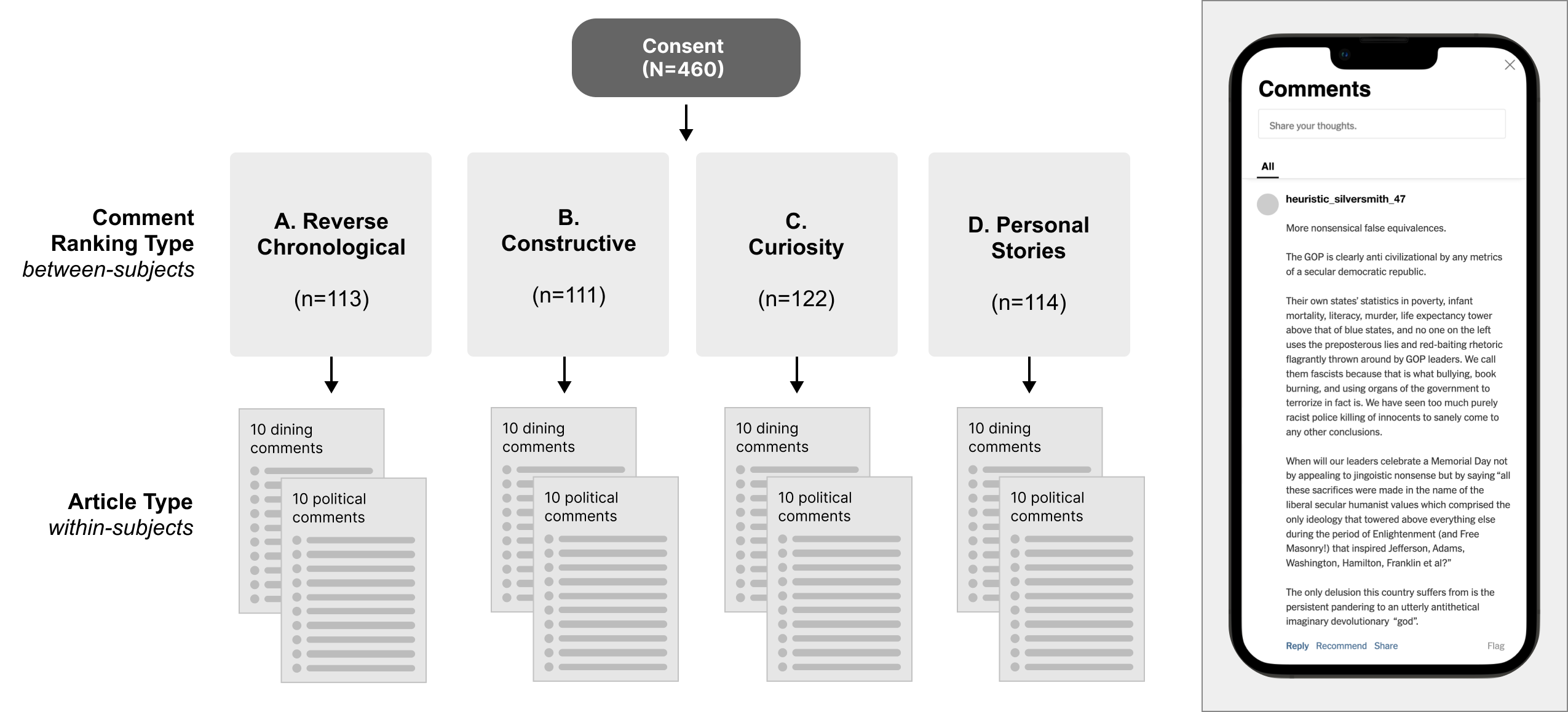}
  \caption{On the left, diagram depicting the breakdown of experimental conditions for participants. The comments from condition A. were taken directly from the displayed order on nytimes.com site as a control. The comments for conditions B., C., and D. were ordered by the top-scored comments for each respective classifier, excluding replies. On the right, example of an interactive prototype accessed by participants in condition A. for the politics article. User profile images, engagement data, New York Times-specific branding and features were removed. User names were randomly generated, with names consistent between conditions. }
  \Description{A top block with the full participant count, N=460, leading four blocks with the condition names and counts, each leading into a panel with two article types displaying 10 strips representing comments. }
\end{figure}

We employed a 4 (Ranking Type: Control, Constructive, Curiosity, Personal Stories) x 2 (Article Type: Dining, Politics) experimental design, as shown in Figure 1. Participants applied randomly to one of four comment ranking conditions, each containing a different subset of 10 comments from the full comments section. All participants saw both article types. The comments selected for control and treatment conditions were from two public New York Times articles: an article in the Dining section, "The 100 Best Restaurants in NYC" \cite{The100Be81:online} with 556 comments, and a Politics Op-Ed: "The Politics of Delusion Have Taken Hold" \cite{OpinionT34:online} with 1317 comments.\footnote{These articles were selected for their contrast in subject matter, and for featuring well-populated comment sections. These comments were not used in model training.}
Participants used a scrollable Figma mobile prototype (Figure 1) to view comments. Prototype stimuli are included in Appendix A. 

During the survey, respondents were first asked about baseline commenting habits, then shown a short article description and header image with brand imagery removed, and given 1-2 minutes to review the first dining comments prototype while thinking aloud and screen recording their usage. They then repeated the process for the politics op-ed, which also included extra questions related to polarization, described in Section 3.3. 

\subsection{Data Analysis}
After viewing each prototype, participants rated the comment sections according to our four types of outcome measures, captured in the eight total DVs. Our outcome categories were inspired by the “building blocks” of flourishing online spaces identified by New\_ Public \cite{CivicSig0:online}. Full question text in Appendix B. 
\subsubsection{Participation.} Perceived likelihood of \textit{a. Hostility} or \textit{b. Respect} for new commenters entering a comments section. These variables aim to gauge how "welcoming" the atmosphere of a comments section is, independent of whether a participant was personally interested or willing to join a conversation, based on individual factors.
\subsubsection{Credibility.} Perceived \textit{c. Trustworthiness} and \textit{d. Informativeness} aim to measure aspects of credibility, mapping to the “Understand” building block \cite{CivicSig0:online}, which includes \textit{“show reliable information.”} 
\subsubsection{Engagement.} Perceived \textit{f. Interestingness}, and likelihood of \textit{e. Future Visit} aim to capture participants' self-reported engagement with the content. Though engaging content might not necessarily be "prosocial," we also wanted to understand these effects given the business considerations for deploying any intervention, it is important to understand if up-ranking “prosocial” content might have tradeoffs in reducing engagement on a platform.
\subsubsection{Polarization.} \textit{g. Democrat Favorability} and \textit{h. Republican Favorability} were measured for the politics article to capture how favorable the participant perceives the other commenters are towards these groups, rather than their own personal feelings. Wording was adapted from the ‘partisan animosity’ measure used in the 2022 Strengthening Democracy Challenge megastudy \cite{voelkel2023megastudy}. This outcome also maps to the “Connect” building block \cite{CivicSig0:online} to “build bridges between groups.” 

We first conducted two-way ANOVAs for each DV, followed by one-way ANOVAs\footnote{Given that we were interested in the effects of the Ranking Type independently within each Article Type, and expected the article context (dining or politics) to fundamentally alter the interpretation of DVs, we chose to examine these effects separately in separate one-way ANOVAs and post-hoc tests for each article.} for each DV by article, with post-hoc tests using the Tukey HSD method to make pairwise comparisons between the four ranking conditions within the politics and dining conditions. We also asked several open-response responses, for example to explain which specific comments they found most and least valuable, and why. The prototype think-aloud videos were automatically transcribed by the dscout software and analyzed as text alongside the open-response entries. We used exploratory coding to identify initial themes\footnote{Specifically, we did two rounds of open-coding with members of the team for a sample of 100 entries across conditions, selected chronologically, followed by discussion and comparison of themes to create a draft codebook.} in order to shed further light on findings in Section 5.

\section{Results}
\subsection{Impact of Article and Ranking Type on DVs}
A series of two-way ANOVAs found that the Article condition had statistically significant effects on all DVs, which indicates that perceptions of the prosocial qualities of a comment section are dependent on the context of the corresponding article. We also found that the Ranking condition had statistically significant effects on all DVs, except for \textit{Hostility}, which suggests that the Ranking condition does indeed have an influence on most DVs across articles. There were also significant interaction effects, which means that both Article and Ranking Type independently and interactively affect how users perceive the prosociality of a comments section. For more detail, see Figure 5 in Appendix D.

\begin{figure}[h]
  \centering
  \includegraphics[width=\linewidth]{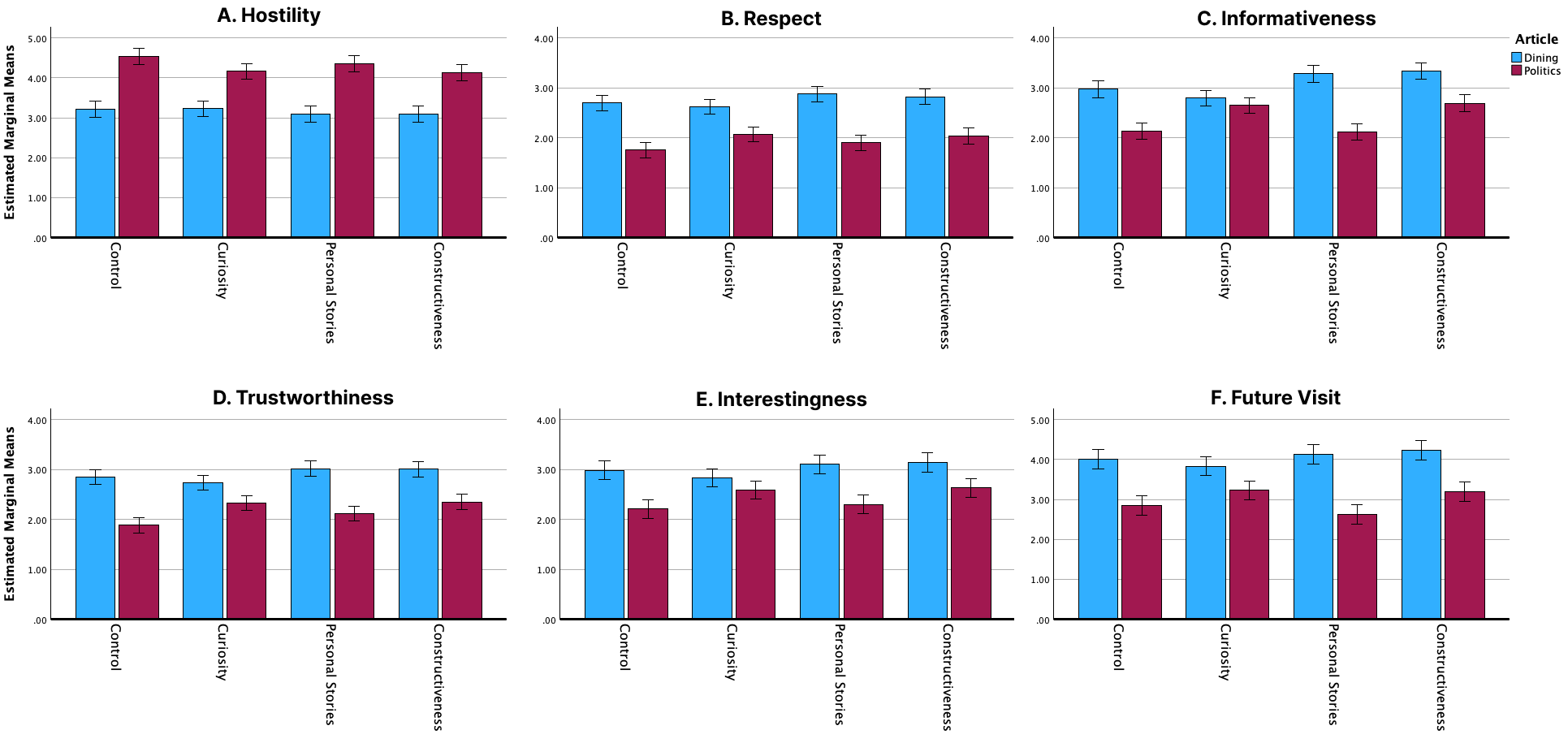}
  \caption{The means of \textit{Hostility} (a), \textit{Respect} (b), \textit{Informativeness} (c), \textit{Trustworthiness} (d), \textit{Interestingness} (e), and \textit{Future Visit} (d) in the dining article (blue), and politics article (red) for each ranking condition, shown with 95\% Confidence Interval error bars.}
  \Description{}
\end{figure}

\subsection{Comparisons between Ranking Conditions in the Politics Article}
Post-hoc comparisons using the Tukey HSD test revealed a number of significant effects of the constructiveness and curiosity conditions on a number of the DVs. There were no significant differences between the personal stories and control conditions across DVs.
\subsubsection{Participation.} The mean value of \textit{Hostility} was significantly lower in the curiosity condition (M = 4.16, SD = 1.13), p = .03, and the constructiveness condition (M = 4.135, SD = 1.065), p = .02 compared to the control condition (M = 4.53, SD = .90). There was no significant difference between control and personal stories (p = .58). For \textit{Respect}, compared to control (M = 1.752, SD = .94), the mean value was significantly higher in the curiosity condition (M = 2.07, SD = .93), p = .38  but not in the constructiveness (p = .08) or personal stories (p = .58) conditions.
\subsubsection{Credibility.} The mean value of \textit{Informativeness} was also significantly higher in the curiosity condition (M =2.65, SD = .97), p = <.001, and the constructiveness condition (M = 2.69, SD = .92), p = <.001 compared to the control condition (M = 2.13, SD = 1.05). There was no significant difference between control and personal stories (p = 1). \textit{Trustworthiness} followed the same pattern; compared to control (M = 1.89, SD = .93), the mean value was significantly higher in the curiosity condition (M = 2.33, SD = .94), p = .001  and in the constructiveness condition (M = 2.35, SD = .86), p = <.001, but not the personal stories condition (p = .23).
\subsubsection{Engagement.} The mean value of \textit{Interestingness} was significantly higher in the curiosity condition (M = 2.59, SD = 1.13), p = .049, and constructiveness condition (M =2.63, SD = 1.15), p = .027 compared to control (M = 2.21, SD = 1.11), but not in personal stories (p = .939). There was no significant difference in the mean values of \textit{Future Visit} across conditions. 
\subsubsection{Polarization.} The mean was significantly higher for \textit{Republican Favorability} in the constructiveness condition (M = 2.62, SD = .97), p = .033, compared to control (M = 2.24, SD = 1.10). There were no statistically significant differences in means for \textit{Democrat Favorability} in curiosity (p = .136), personal stories (p = .170), or constructiveness (p = .268) compared to control. 
\begin{figure}[h]
  \centering
  \includegraphics[width=\textwidth]{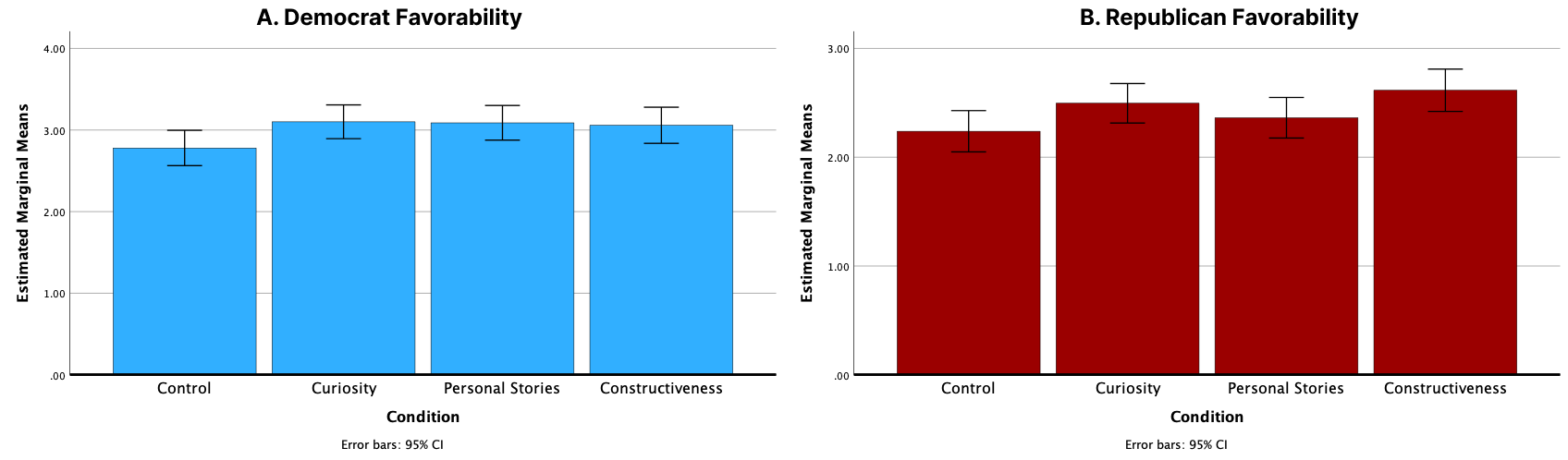}
  \caption{The means of \textit{Democrat Favorability} (a) and \textit{Republican Favorability} (b) in the comment section by each ranking condition, on a scale from "Very Unfavorable" (1) to "Very Favorable" (5). Question wording: \textit{"In the previous comment section, how would you rate the commenters' attitudes towards [Republican or Democratic] voters?"}}
  \Description{}
\end{figure}

\subsection{Comparisons between Ranking Conditions in the Dining Article}
\textit{Informativeness} was found to have a significant difference from control (M = 2.98, SD = .81), in the constructiveness condition (M = 3.33, SD = .68), p = .004, and personal Stories condition (M = 2.88, SD = .71), p = .018, but not in the curiosity condition (p = .87). The ranking type didn’t lead to any other significant differences in mean value from the control condition for the other DVs, \textit{Hostility}, \textit{Respect}, \textit{Trustworthiness}, \textit{Interestingness}, or \textit{Future Visit}, for the dining article. Polarization outcomes were not measured.

\section{Qualitative results and Discussion}
We found that most significant changes, across the categories of \textit{Participation}, \textit{Credibility}, \textit{Engagement}, and \textit{Polarization}, occurred in the politics article, and particularly in the curiosity and constructiveness conditions, described as not containing \textit{“any comments that [they] thought were very bad,”} in contrast to the control condition, where one participant noted \textit{“[none of] them were valuable.”} Although the dining article showed less of an effect across DVs, the personal stories and constructiveness conditions were perceived as significantly more informative compared to control. 

Overall, both the significant interaction effects between article and ranking conditions, as well as the differences in the significant effects within ranking conditions for each article, make clear that the specific subject-matter of a comments section has a drastic effect on how “prosocial” it is perceived as. This suggests that any prosocial attribute, or combination of attributes, used to re-rank comments must be evaluated independently, as what is prosocial for one context might not be prosocial for another context.

There were several other interesting findings when honing in on the specific effects between, and across, ranking conditions for each article:
\begin{itemize}
\item \textbf{“[With personal stories,] I’m not usually spending that much time looking for one-off anecdotes.”} Personal stories proved to be especially sensitive to article context. As shown in Section 4.3, the condition was significantly higher than control in \textit{Informativeness} for the dining article. Participants resonated with \textit{“true experiences people have had … and maybe [gaining] more information on a specific restaurant.”} At the same time, personal stories had no significant effects in the politics condition. In the context of a political discussion, such stories were interpreted as containing \textit{“a lot of emotion”} or \textit{“bias.”} Future work might address this by exploring layering in other attributes, for example ensuring a personal story is also constructive or curious.
\item \textbf{“These [curiosity comments] seem like rhetorical questions that I wouldn't respond to or even interact with.”} The effects of curiosity also varied greatly between articles. For the politics article, it resulted in significant prosocial outcomes across most measures (with constructiveness close behind). Yet in dining, Section 4.3 showed it was rated as significantly less informative. Qualitatively, participants reading the dining article comments were more interested in recommendations and felt that in curiosity that \textit{“nobody [was] really giving any information”}. In contrast, curiosity was more appreciated in politics, with one saying \textit{“this is impressive…these are readers who deeply, critically think about what they’re reading.”} This indicates that curiosity might be especially valuable in sensemaking contexts, like political debate, where questioning expands thought, rather than functional contexts like restaurant reviews, where users are seeking specific information.
\item \textbf{“Immediately the longer ones [in constructiveness] are kind of a turnoff for me.”} Stylistic qualities may have also had an effect on the participants’ reported outcomes, such as \textit{Future Visit}. Responses revealed a strong aversion to overly lengthy comments presented across conditions, and especially in the constructiveness condition, where the comment length was typically longer than any other condition’s comments. Some found these \textit{“very thought out”} but many others found it \textit{“a little exhausting to read it all.”} 
\item \textbf{“[The curious and constructive] comments are valuable in terms of understanding where people are coming from.”} Finally, this work shows preliminary support that re-ranking by constructiveness and curiosity, but not personal stories, can have some influence on partisan animosity, if not significantly so. Participants in the constructiveness condition, for example, noted that they didn’t seem to contain as much \textit{“hate and racism”} as the political comments sections they were used to seeing. This was in contrast to the control condition, reflecting the current status quo of a news comment section, even after the moderation needed for them to appear on the site, were still described by one as \textit{“heavily biased”} and \textit{“unreasonably hateful.”}
\end{itemize}
In sum, these findings show the potential of re-ranking by prosocial qualities to influence users across a number of meaningful social outcomes such as \textit{Participation}, \textit{Credibility}, and \textit{Polarization}, particularly in fraught political discussions. If other factors such as verbosity and subject-matter context are taken into account, we also believe this can be done without sacrificing \textit{Engagement.}

\section{Limitations and Future Work}
We designed the prototypes to most closely simulate user behavior on a news site, such as scrolling. However, we did not simulate full features such as news source, engagement data, and threaded comments. In real life these features may influence how a user interacts with the comments sections. For our study, these were hidden to better isolate the effect of the comment ranking alone. Future studies might include these features via fuller-featured prototypes, extensions, or on-platform testing to better measure real-world behaviors and outcomes. 

Relatedly, we focused only on two predetermined topics and article comment sections, which we limited to 10 comments per section without reply threading. Future studies could explore methods using comments from a users’ own comment ecosystem, such as diary studies or user-donated data. 
Additionally, this study explores only three attributes in the context of one particular platform’s comment section. Future work should continue to compare additional attributes of language and evaluate their performance across different linguistic and social contexts for a variety of social outcomes. In particular, we recommend study of polarization variables beyond partisan animosity, such as partisan violence \cite{voelkel2023megastudy}, and factoring participants’ political affiliations into recruitment. 

In future work, we recommend continuing to evaluate a variety of interventions made possible with prosocial classifiers, including re-ranking mixed with engagement signals, layering attributes (for example, personal stories and constructiveness), enabling sorting and filtering by attributes in explicit user controls, and nudges or rewards to comment writers based on their comment.
\section{Conclusion}
This paper assessed the effects of up-ranking comments by comparing three treatment conditions — constructiveness, curiosity, and personal stories — to the control condition of reverse-chronological rankings. Our results show significant prosocial improvements along multiple dimensions such as \textit{Respect}, \textit{Informativeness}, \textit{Trustworthiness}, and \textit{Interestingness} when ranking comments by constructiveness and curiosity in politics articles. We also saw that results were sensitive to article context, as the dining article revealed different patterns of effects, with only \textit{Informativeness} improving significantly in the constructiveness and personal stories conditions. Personal stories did not impact any other the prosocial attributes in the politics or dining article, though was perceived as qualitatively more valuable in dining, and occasionally valuable in politics. Overall, we demonstrate the promise of content-based ranking signals to promote prosocial outcomes in online comment sections, especially as it pertains to politically divisive  topics. 

\section{Acknowledgments}

Huge thanks to all who contributed by reviews of this research, with special thanks to Rachel Xu, Beth Goldberg, Lucas Dixon, and Greg Holyk, Alyssa Chvasta, Daniel Borkan, Thea Mann, and Jeffrey Sorenson. Thank you as well to participants in the study for sharing their experiences about online comment sections.

\bibliographystyle{ACM-Reference-Format}
\bibliography{sample-base}

\appendix

\section{Prototype Stimuli}
All eight prototype are linked below, and can be accessed by entering the password "test."
\begin{enumerate}
    \item \href{https://www.figma.com/proto/qaeyO6xJtdu3682mnRtfPw/Comments-stimuli?page-id=0\%3A1\&type=design\&node-id=262-431\&viewport=-1342\%2C-970\%2C0.59\&t=WGSxc1Yc9A5GmfYd-1\&scaling=scale-down}{Control Ranking, Dining Article}
    \item \href{https://www.figma.com/proto/qaeyO6xJtdu3682mnRtfPw/Comments-stimuli?page-id=0%3A1&node-id=262-435&scaling=scale-down&t=RUXlYjvZY4DQp00y-1}{Constructiveness Ranking, Dining Article}
    \item \href{https://www.figma.com/proto/qaeyO6xJtdu3682mnRtfPw/Comments-stimuli?page-id=0%3A1&type=design&node-id=262-444&viewport=-3517%2C-961%2C0.53&t=MPZZYKvKzHojBtPZ-1&scaling=scale-down}{Curiosity Ranking, Dining Article}
    \item \href{https://www.figma.com/proto/qaeyO6xJtdu3682mnRtfPw/Comments-stimuli?page-id=0%3A1&type=design&node-id=262-440&viewport=-943%2C-96%2C0.26&t=SxPstaQlbUzBF01v-1&scaling=scale-down}{Personal Stories Ranking, Dining Article}
    \item \href{https://www.figma.com/proto/qaeyO6xJtdu3682mnRtfPw/Comments-stimuli?page-id=0%3A1&type=design&node-id=262-434&viewport=-2011%2C-1181%2C0.64&t=oIgkqe2hPKrnq7ta-1&scaling=scale-down}{Control Ranking, Politics Article}
    \item \href{https://www.figma.com/proto/qaeyO6xJtdu3682mnRtfPw/Comments-stimuli?page-id=0%3A1&type=design&node-id=262-438&viewport=-1934%2C-561%2C0.47&t=bmOMCMXrv3wnJwnn-1&scaling=scale-down}{Constructiveness Ranking, Politics Article}
    \item \href{https://www.figma.com/proto/qaeyO6xJtdu3682mnRtfPw/Comments-stimuli?page-id=0%3A1&type=design&node-id=262-445&viewport=-3517%2C-961%2C0.53&t=Nz1LApkC4UcYPH3V-1&scaling=scale-down&mode=design}{Curiosity Ranking, Politics Article}
    \item \href{https://www.figma.com/proto/qaeyO6xJtdu3682mnRtfPw/Comments-stimuli?page-id=0%3A1&type=design&node-id=262-443&viewport=-2366%2C-501%2C0.43&t=VI2uWAtQpwFubyJx-1&scaling=scale-down}{Personal Stories Ranking, Politics Article}
\end{enumerate}

\section{Survey Text}
The original participant screening survey and main survey text is available as a supplementary material.

\begin{itemize}
    \item \textit{Hostility}. [Multiple Choice, SINGLE SELECT] "How likely do you think it is that a new commenter in this comment section would receive hostile replies?" (Very likely, Somewhat likely, Neither likely or unlikely, Somewhat unlikely, Very unlikely)
    \item \textit{Respect}. [Multiple Choice, SINGLE SELECT] "If you were to leave a new comment on this article, how confident are you that others commenters would treat you with respect?" (Very confident, Somewhat confident, Not very confident, Not at all confident)
    \item \textit{Informativeness}. [Multiple Choice, SINGLE SELECT] "How informative is the comment section on this topic?" (Very informative, Somewhat informative, Not very informative, Not at all informative)
    \item \textit{Trustworthiness}. [Multiple Choice, SINGLE SELECT] "How trustworthy or untrustworthy do you find the information in this comment section?" (Very trustworthy, Somewhat trustworthy, Somewhat untrustworthy, Not at all trustworthy)
    \item \textit{Interestingness}. [Multiple Choice, SINGLE SELECT] "How interesting or uninteresting are these comments to you, personally?" (Very interesting, Somewhat interesting, Somewhat uninteresting, Not at all interesting)
    \item \textit{Interestingness}. [Multiple Choice, SINGLE SELECT] "How likely would you be to view the comments on a similar article on this site in the future?" (Very likely, Somewhat likely, Neither likely nor unlikely, Somewhat unlikely, Very unlikely)
    \item \textit{Democrat Favorability}. [Multiple Choice, SINGLE SELECT] "In the previous comment section, how would you rate the commenters' attitudes towards *Democratic voters*?" (Very unfavorable, Somewhat unfavorable, Neither favorable nor unfavorable, Somewhat favorable, Very favorable)
    \item \textit{Republican Favorability}. [Multiple Choice, SINGLE SELECT] "In the previous comment section, how would you rate the commenters' attitudes towards *Republican voters*?" (Very unfavorable, Somewhat unfavorable, Neither favorable nor unfavorable, Somewhat favorable, Very favorable)
\end{itemize}

\section{Model Card and Statistical Tests}

This section contains the Model Card for the classifiers used in the study. It also contains tables generated in SPSS with the Two-way ANOVA Tests of Between-Subjects Effects across conditions, the Tukey HSD post-hoc tests of a one-way ANOVA for the Dining article data, and the Tukey HSD post-hoc tests of a one-way ANOVA for the Politics article data.

\begin{figure}[h]
    \centering
    \includegraphics[width=1\linewidth]{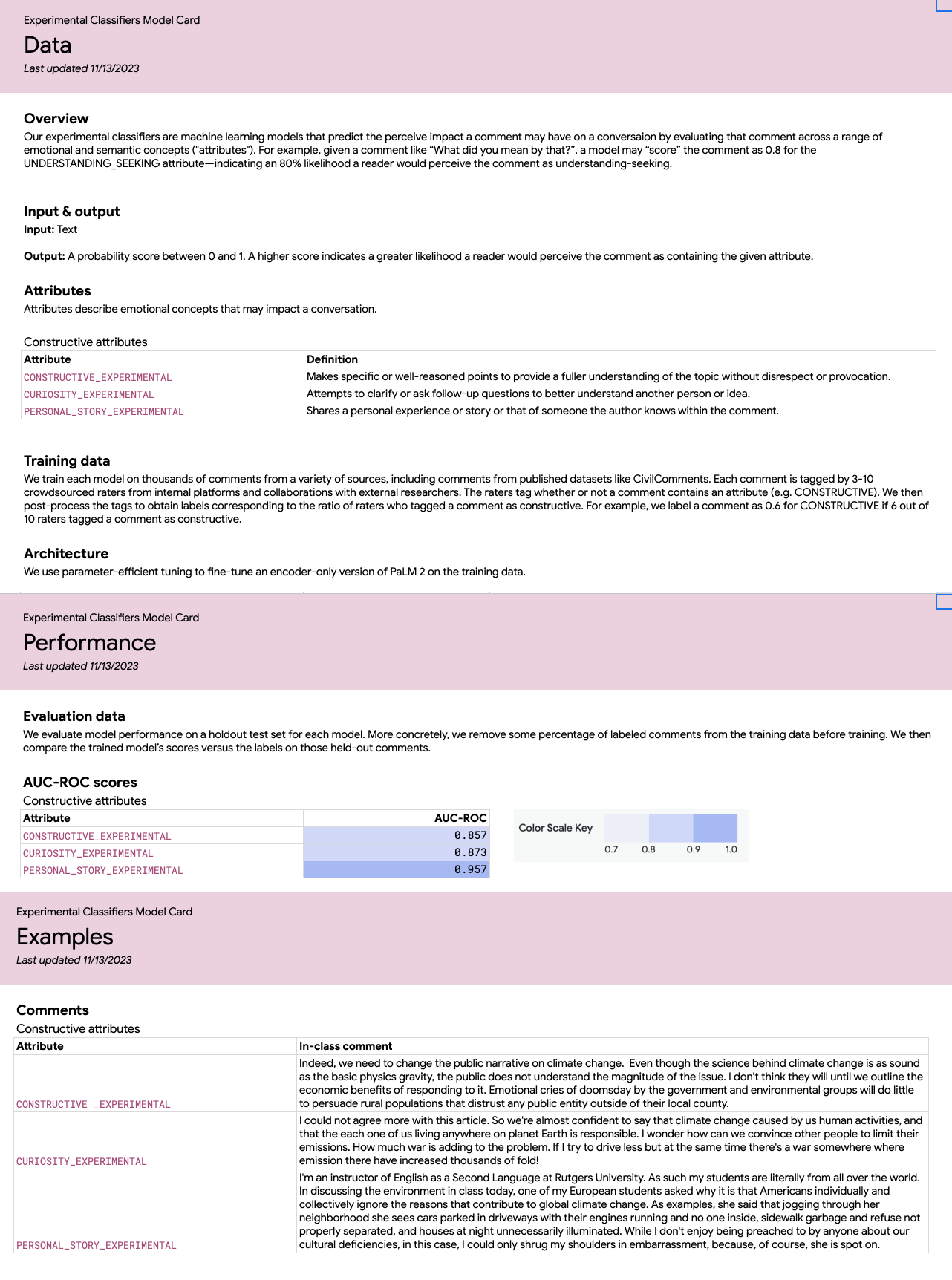}
    \caption{The model card for constructiveness, curiosity, and personal stories.}
    \label{fig:enter-label}
\end{figure}

\begin{figure}[h]
    \centering
    \includegraphics[width=.9\linewidth]{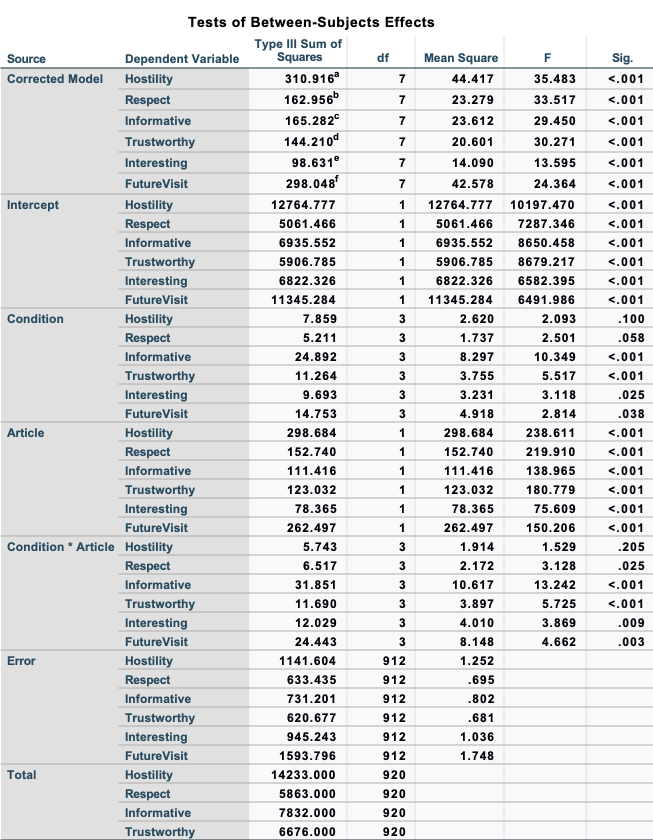}
    \caption{Two-way ANOVA Tests of Between-Subjects Effects}
    \label{fig:enter-label}
\end{figure}

\begin{figure}
    \centering
    \includegraphics[width=0.55\linewidth]{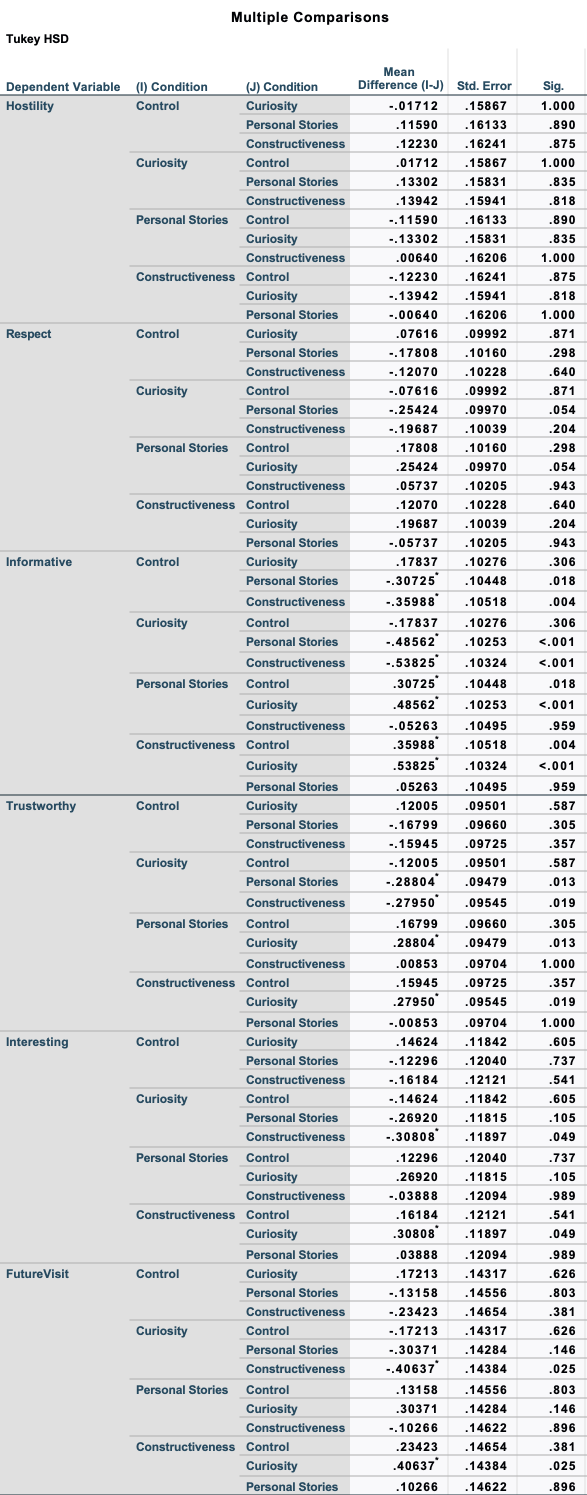}
    \caption{Tukey HSD post-hoc tests of a one-way ANOVA for the Dining article data.}
    \label{fig:enter-label}
\end{figure}

\begin{figure}
    \centering
    \includegraphics[width=.4\linewidth]{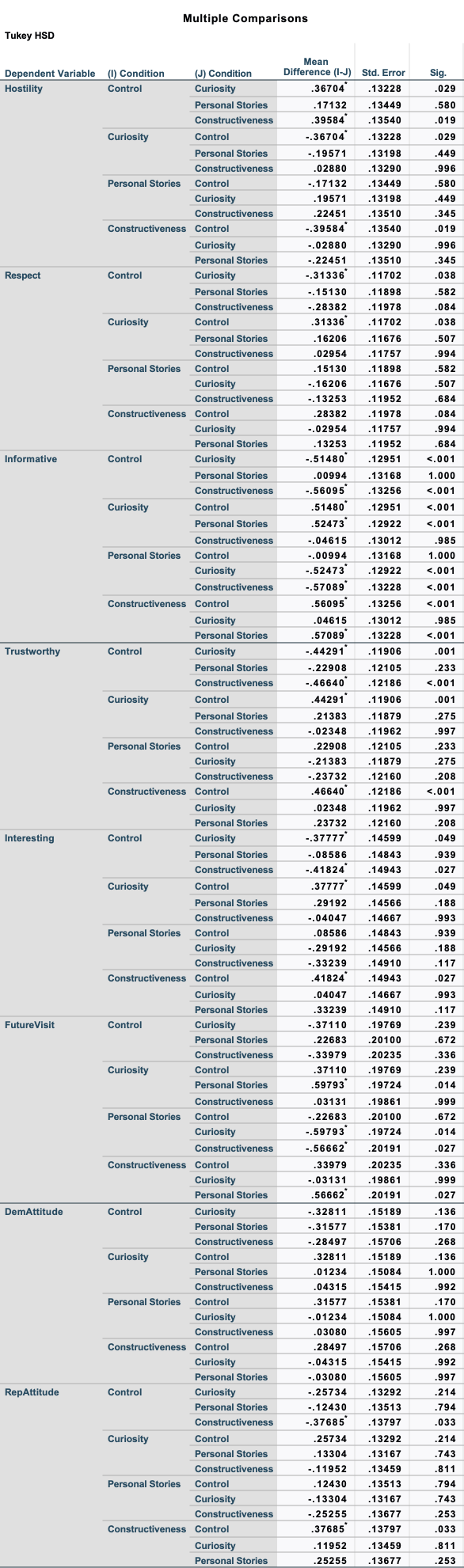}
    \caption{Tukey HSD post-hoc tests of a one-way ANOVA for the Politics article data.}
    \label{fig:enter-label}
\end{figure}

\end{document}